 \documentclass[preprint2]{aastex}
  \usepackage{psfig}

\shorttitle{W44 Pulsar Wind Nebula}
\shortauthors{Petre, Kuntz, \& Shelton}

\begin{document}
\title{The X-ray Structure and Spectrum of the Pulsar Wind
Nebula Surrounding PSR B1853+01 in W44}
\author{R. Petre, K. D. Kuntz \altaffilmark{1}}
\affil{NASA Goddard Space Flight Center, Greenbelt, MD 20771}
\and \author{R. L. Shelton}
\affil{The Henry A. Rowland Department of Physics and Astronomy,
Johns Hopkins University, 3400 North Charles Street, Baltimore, MD 21218}
\altaffiltext{1}{Department of Physics, University of Maryland 
Baltimore County,
1000 Hilltop Circle, Baltimore MD 21250}

\begin{abstract}

We present the result of a Chandra ACIS observation of the pulsar
PSR~B1853+01 and its associated pulsar wind nebula (PWN), embedded within
the supernova remnant W44. A hard band ACIS map cleanly distinguishes
the PWN from the thermal emission of W44. The nebula is extended in
the north-south direction, with an extent about half that of the radio
emission. Morphological differences between the X-ray and radio
images are apparent. Spectral fitting reveals a clear difference
in spectral index between the hard emission from PSR~B1853+01
($\Gamma \sim$ 1.4) and the extended nebula ($\Gamma \sim$ 2.2).
The more accurate values for the X-ray flux and spectral index are
used refine estimates for PWN parameters, including magnetic field 
strength, the average Lorentz factor $\gamma$ of the particles in the wind, the 
magnetization parameter $\sigma$, and the ratio k of electrons to 
other particles.

\end{abstract}

\section{Introduction}

The remarkable pulsar wind nebula associated with PSR~B1853+01 and 
embedded in the evolved, mixed-morphology supernova remnant W44 is of 
particular interest for several reasons. It is the oldest known 
pulsar wind nebula (PWN), associated with an active pulsar (only the
PWN in IC~443 is thought to be older -- Bocchino \& Bychkov 2001).  Its
age of 20,000 yr is estimated from the 
PSR~B1853+01 spindown (Wolszczan, Cordes \& Dewey 1991). At the same 
time, PSR~B1853+01 is one of the 10 youngest known pulsars. Thus the 
nebula
may allow the testing of hypotheses regarding PWN evolution,
and serves as a bridge between the young, active pulsars in supernova 
remnants
and the preponderant population of isolated, old radio pulsars.
Additionally, because of its high proper motion, the pulsar leaves a record
of its evolution embedded in the extended nebula. A
combination of high resolution radio and X-ray observations can
potentially disentangle this record. For example, radio measurements of 
the
extent of the nebula perpendicular
to the direction of motion provide means for setting an upper limit
to the lifetime of the radio emitting electrons ($\sim$ 6,000 yr) 
unavailable from 
observation of wind nebulae associated with stationary pulsars.  Finally, some of the properties of this PWN are 
similar enough to those of the much younger and more luminous Crab 
Nebula (Chevalier 2000) to invite comparisons and speculation 
regarding the reason for the similarities.

With the clearest view of the X-ray universe, especially above 3 keV, 
now available via the Chandra X-ray Observatory, more comprehensive 
studies of embedded PWNs become feasible. In this paper we use 
Chandra's Advanced CCD Imaging Spectrometer (ACIS) to reveal the X-ray
structure and spectrum of the synchrotron nebula surrounding the W44
pulsar, and provide a more careful look at the X-ray spectrum and
its spatial variation. As has been shown in numerous other works
(e.g., Harrus et al. 1996; Frail et al. 1996; Torii et al. 2000),
the study of the wind nebulae surrounding pulsars provides a means
for understanding the energetics of pulsars, and in particular how
they transfer their rotational spindown energy into a relativistic
wind.

W44 is one of the
first remnants for which hard band X-ray imaging was used to isolate a
pulsar and its associated PWN from the brighter, softer
thermal X-ray emission associated with the remnant's shock-heated 
gas. Neither the pulsar nor the nebula is apparent in low energy 
X-ray images, such as that from the ROSAT PSPC (Rho et al. 1994). 
Using ASCA and its broader band imaging, however,
Harrus et al. (1996) showed that while the
PWN is invisible in the broad band image, it becomes the
dominant feature above 4 keV. The centroid of the X-ray emission is 
consistent
with the location of the pulsar, PSR~B1853+01 (Wolszczan, Cordes \&
Dewey 1991). ASCA's modest angular resolution precluded spatially 
distinguishing the synchrotron nebula from the surrounding
diffuse emission, but Harrus et al. showed that the spectrum of the
region including the PWN has a hard continuum component not detected
elsewhere in W44. The techniques pioneered in Harrus et al. (1996) have
subsequently been
used to identify stellar remnants or synchrotron nebulae
in other remnants (e.g., IC~443 - Keohane et al. 1997;
MSH~15-56 - Plucinsky 1998; G292.0+1.8 - Torii, Tsunemi, \& Slane
1998). The most important consequence of this approach is the
dramatic increase in the number of supernova remnants with identified
synchrotron nebulae and/or compact stellar remnants.

The discovery of the X-ray synchrotron nebula occurred
contemporaneously with the mapping of the pulsar wind nebula in the
radio (Frail et al. 1996). At
1.4 GHz it appears cometary in shape with an extent of $\sim$2.5 arc minutes.
The pulsar is located at the southern extremity. The radio surface
brightness peaks at the widest part of the tail, approximately 1 arc minute
north of the pulsar. Frail et al. interpret this structure as the
result of the pulsar's motion through the interior of the remnant.
Using three independent techniques, they derive a velocity
of the pulsar of approximately 375 km/s. The radio emission has a 
nonthermal spectrum with a spectral index of -0.12$\pm$0.04, and it 
is 17$\pm$4 percent
polarized. The spectral index distinguishes the PWN from
the surrounding emission associated with W44 ($\Gamma\sim$-0.33);
the spectrum and degree
of polarization are similar to other pulsar wind nebulae.
Using a combination of the X-ray and
radio properties, Frail et al. estimate some key pulsar wind nebula
parameters, including magnetic field strength, and the Lorentz factor
$\gamma$ of the
electrons near the spectral break between the radio and X-ray slopes.

Giacani et al. (1997) presented radio and X-ray images of W44 as a
whole. The PWN is apparent but inconspicuous in the radio. A line
of H$\alpha$ filaments lies along the eastern edge of the PWN, but there
is no general correspondence with radio features, and it is unclear
whether this emission is associated with the PWN or with shock heated
material near the PWN only in projection.

The distance to W44 has generally been taken to be around 3 kpc, 
based on measurements of H~I absorption and 1720 MHz maser lines 
(Caswell et al. 1975; Claussen et al. 1996).  The analytical modeling 
of Cox et al. (1999) refines this distance to be between 2.5 and 2.6 
kpc.  We use a distance of 2.6 kpc, and scale parameters in 
terms of d$_{2.6}$.  The implications of using this refined value are 
minor.  A luminosity estimate, for instance, is reduced by 25 
percent, which is probably well within the uncertainty of the 
estimate.

\section{Observations and analysis}

W44 was observed using ACIS-S on 31 October, 2000 for 45.5 ks. As
the primary goal of the observation was to understand the nature of
the centrally peaked thermal emission in this mixed-morphology SNR,
the S3 chip was pointed at the remnant center. The results of that
investigation will be reported elsewhere. A fortuitous observation 
date made it possible to orient the spacecraft roll angle to place
the pulsar squarely within the front-illuminated S2 chip.
Although the calibration of S2 is not as complete as that of S3, it is
adequate for the analysis we perform here.

  The data were analyzed using a combination of public an
d custom
software tools.  We used CIAO tools version
2.2 and the calibrations available with the CIAO calibration database
2.9. An initial image of the pulsar nebula in the 2.0-8.0 keV band
was formed from the events file after point source removal, and
smoothed with a $4\farcs4$ (9 ACIS pixel) HWHM Gaussian.
Spectra were extracted using custom IDL software designed to extract counts
within predefined brightness contours.
Spectral fitting was performed using XSPEC v.11.1.0.

\section{Images}

\begin{figure*}
\vspace*{-0.5cm}
\centerline{\psfig{figure=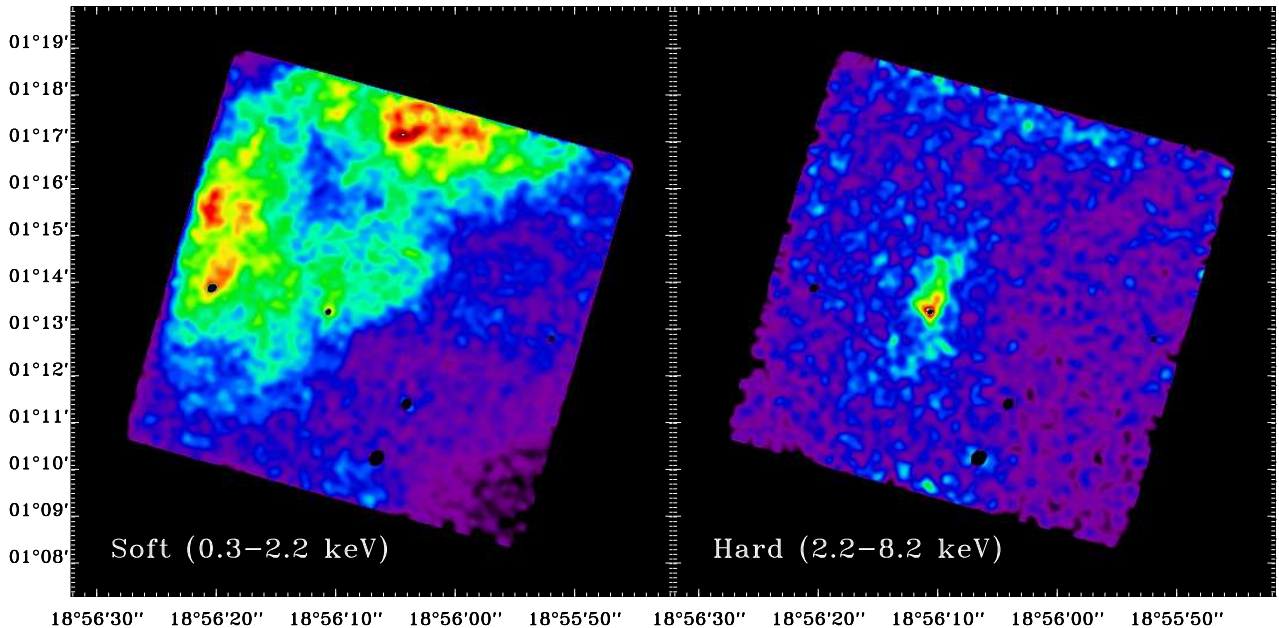,width=16.0cm,angle=90.0}}
\vspace*{0.5cm}
\caption{{\it a)} The soft (0.3-2.2 keV) broadband image from the ACIS-S2 chip.
The black ellipses are the regions containing point sources
which were removed before smoothing by a $4\farcs4$ HWHM Gaussian.
The ellipses enclose 95\% of the point source flux,
and show the shape and orientation of the PSF.
{\it b)} The hard (2.2-8.2 keV) broadband image from the ACIS-S2 chip.
The image has been smoothed by a $4\farcs4$ HWHM Gaussian.}
\label{fig:images}
\end{figure*}

In Figure~\ref{fig:images}
we show images of the region of W44 in the S2 field of
view. The soft band (0.3-2.2~keV) image (Fig. 1a) is dominated by 
diffuse thermal
emission from W44. Several unresolved sources are apparent; the
pulsar is one of these, though not the brightest. The pulsar is
located along the edge of an emission plateau stretching approximately
southeast to northwest. It is the relatively low brightness of the
pulsar and its fortuitous location along the ridge that rendered it
undetectable to the ROSAT HRI, whose 0.2-2.4 keV band pass is similar 
to that of the soft band image displayed here. In contrast, the 
surface brightness of the thermal emission from W44 is substantially 
lower in the hard band (2.2-8.2~keV) image
(Fig. 1b).
The unresolved sources are all still apparent; again, the pulsar is
not the brightest. What makes the pulsar stand out is the enveloping
diffuse emission, surrounding it and extending northward.

Using the hard image to avoid contamination by diffuse emission, 
we extracted the following position for 
PSR~B1853+01: 18h~56m~10.653$\pm$0.028s; 01$\arcdeg$~13$\arcmin$~21.3$\pm$0.36$\arcsec$ (J2000).
This can be compared with the radio position 18h~56m~10.8s;
01$\arcdeg$~13$\arcmin$~28$\arcsec$ (Frail et al. 1996).  
The origin of the 7$\arcsec$ offset in declination is unknown.  Frail
et al. (1996) assign an error of 1$\arcsec$.5 to the radio position.
It is larger than the nominal 3$\arcsec$ X-ray coordinate uncertainty
from Chandra (Chandra Observatory Proposer's Guide), although the
uncertainty for an outlying chip might be larger than this.
Interestingly, when undertainties are taken into account, 
the X-ray position is consistent with the presumably
less accurate position from radio timing reported by Wolszczan, Cordes
\& Dewey (1991) at 18h~56m~10.9s;
01$\arcdeg$~13$\arcmin$~20$\arcsec$.6.  
It should be noted that the offset cannot be a result of
proper motion, which has been estimated to be 25 mas/yr (Frail et al. 1996).

\begin{figure*}
\centerline{\psfig{figure=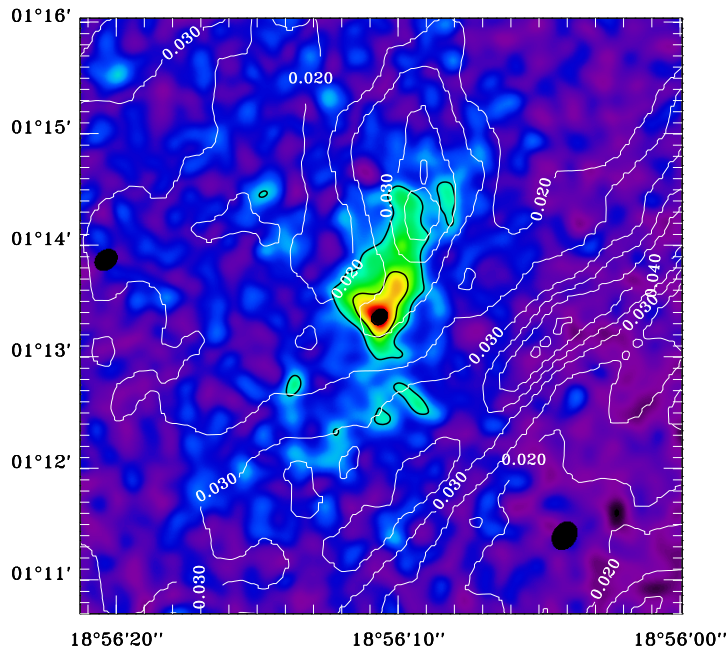,width=8.0cm,angle=90.0}}
\vspace*{0.5cm}
\caption{The region surrounding the PWN.
{\it Grayscale: } The {\it Chandra} image in the 2.2-8.2 band
smoothed with an $4\farcs4$ HWHM Gaussian.
{\it Contours: } The 20cm radio map from Frail et al. (1996),
the units are Jy/beam, where the beam is $15\farcs8\times15\farcs4$.
The dark contours are the boundaries of the subregions used for
spectral fitting.  The black ellipses indicate regions contaminated by point sources.
\label{fig:detail}}
\end{figure*}

In Figure~\ref{fig:detail}
we show a close up of the hard diffuse emission surrounding PSR~B1853+01, 
and compare it with the radio map of Frail et al (1996). Both show
diffuse emission, cometary in shape, trailing toward the north.  The
resemblance is close enough to suggest that this feature is the X-ray
PWN. Nevertheless, clear differences
exist between the X-ray and radio images. Unlike the radio emission, 
the X-ray emission
peaks at the pulsar. Additionally, the X-ray extent is about
half the radio. The real X-ray extent is slightly (about 20 percent) 
less than what is apparent in the figures.  Due to the lack of 
counts, the image has been
smoothed, thus increasing the apparent X-ray extent.

An elliptical region of intermediate X-ray
surface brightness extends 1-2$\arcmin$ in all directions beyond the radio
nebula, and a brighter ridge extends to the south and north. Our
data lack sufficient signal to allow us to determine whether this
emission is nonthermal, and thus associated with the PWN, or is thermal
emission from W44.

\section{Spectra}

We have extracted spectra for the pulsar, the entire PWN region, and
the two subregions indicated in Figure~\ref{fig:detail}.
Using the smoothed image to
create image masks, we extracted spectra from the events file for
regions of the pulsar nebula with surfce brightness between 3.3 and
5.7$\times$10$^{-4}$ counts s$^{-1}$ arcsec$^{-2}$, and surface brightness
greater than 5.7$\times$10$^{-4}$ counts s$^{-1}$ arcsec$^{-2}$ (but
excluding the pulsar itself). We used
count-weighted response files (created with the CIAO routines {\it mkwarf} 
and {\it mkrmf}) to fit the spectra between 2.2 and 8.2 keV. We used
XSPEC to map $\chi^2$ surfaces, and applied the Lampton, Margon, \&
Bowyer (1976) criteria to find the 90\% confidence interval for
the value of the flux from the deabsorbed power-law models.

Great care has been exercised in fitting to remove the contribution
of the thermal emission. Despite Chandra's ability to resolve
spatially the PWN and to reduce the overall level of contamination due
to thermal emission from W44, there is still a substantial thermal
contribution from the foreground and background thermal emission.
During spectral fitting, we used several different fields drawn
from nearby regions to represent the foreground and background 
thermal contribution. Incorporating these into the fits as a 
"background" file left
insufficient signal for fitting the spectra below 2 keV. Thus all
the fits were restricted to a band corresponding roughly to that of 
the hard band image in Fig. 1b. These restricted fits are insensitive
to absorption by the column density of intervening material,
which for W44 is on the order of
5x10$^{22}$~cm$^{-2}$ (Rho et al. 1994). The fit results are 
insensitive to the specific
choice of background field, largely as a result of the low overall
signal-to-noise ratio of the spectra.

We show pertinent quantities from fits to absorbed power-law and
absorbed Bremsstrahlung models in Table 1.
We restricted our fits to these simple models, given the small number
of counts. All the models give formally acceptable fits, with
$\chi^2_{\nu}~\sim~$1.
Although neither model is preferred for any of the spectra
based on the quality of the fit, we interpret the spectrum in terms of
the power law, the expected model for synchrotron emission.

 \begin{deluxetable}{lrrcccccr}
 \tablecolumns{9}
 \tabletypesize{\footnotesize}
 \tablecaption{Fit Values for Pulsar and PWN Spectra}
 \label{tab:fits}
 \tablewidth{0pt}
 \tablehead{
 \colhead{Region} &
 \colhead{Counts} &
 \colhead{Area} &
 \colhead{Gamma} &
 \colhead{Red. $\chi^2$} &
 \colhead{Flux\tablenotemark{b}} &
 \colhead{kT} &
 \colhead{Red. $\chi^2$} &
 \colhead{d.f} \\
 \colhead{ } &
 \colhead{2-8 keV} &
 \colhead{pixels\tablenotemark{a}} &
 \colhead{ } &
 \colhead{ } &
 \colhead{erg cm$^{-2}$ s$^{-1}$} &
 \colhead{ } &
 \colhead{ } &
 \colhead{ } }
 \startdata
 PSR   &   94 &   476 & 1.28$\pm$0.48 & 0.68
 & 5.96$^{+1.56}_{-1.40}\times10^{-14}$ & 55.0$\pm$230.0 & 0.67 &  3\\
 Inner &  164 &  1821 & 2.12$\pm$0.40 & 1.08
 & 7.57$^{+1.45}_{-1.39}\times10^{-14}$ &  5.40$\pm$2.94 & 1.16 &  7\\
 Both  &  576 &  9707 & 2.20$\pm$0.22 & 1.10
 & 2.11$^{+0.25}_{-0.24}\times10^{-13}$ &  4.27$\pm$1.10 & 1.11 & 30\\
 Outer &  412 &  7886 & 2.33$\pm$0.29 & 1.12
 & 1.38$^{+0.20}_{-0.20}\times10^{-13}$ &  3.65$\pm$1.05 & 1.09 & 21 \\
 Back  & 1441 & 76335 &               &      &                &      &
 &    \\
 \enddata
 \tablenotetext{a}{Pixels are $0\farcs492\times0\farcs492$, the ``native''
 ACIS pixel size.}
 \tablenotetext{b}{Unabsorbed 2.2-8.2 keV band flux; calculated using
 the power-law model.}
 \end{deluxetable}

The spectrum of PSR B1853+01 is hard, with a photon index
$\Gamma$~=~1.29$\pm$0.45. There is no evidence for a second, softer,
thermal component expected from the pulsar surface; the lower energy
cut off of 2 keV renders this component invisible. The overall
spectrum of the extended nebula is considerably softer, with
$\Gamma$~=~2.2$\pm$0.2. This value is similar to the 
$\Gamma$~$\sim$~2.3
found by Harrus et al. (1996) for the spectrum of the pulsar plus
nebula. This consistency is understandable given that the flux from
the nebula is a factor of three higher than that from the pulsar. This
value of $\Gamma$ is also similar to the average photon index of the
Crab Nebula (2.2 -- Willingale et al. 2001, and references therein).

\begin{figure*}
\centerline{\psfig{figure=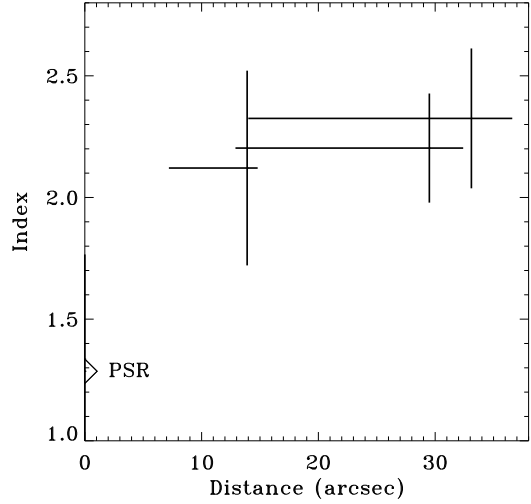,width=8.0cm}}
\caption{The photon index as a function of distance from the pulsar.
The horizontal bars indicate the radii over which the index was
measured. The vertical bars mark the pixel-weighted mean distance
from the pulsar for each region.
\label{fig:index}}
\end{figure*}

In Figure~\ref{fig:index} we show the best-fit photon index
as a function of distance from the pulsar. The best-fit value of $\Gamma$
varies from 2.13 for inner region to 2.33 for the outer region, suggestive
of spectral variation.  The 90 percent confidence range for the 
spectral index from each region, however, contains the best-fit value 
of the other, rendering
the search for spectral variation inconclusive.

\section{Discussion}

The ACIS observation of W44 has revealed the structure and spectrum
of the PWN associated with the W44 pulsar. We find as in previous
work (Harrus et al. 1996) that the pulsar and PWN stand out at
energies above $\sim$2 keV. In the ACIS data they are detected for the
first time at all energies down to the $\sim$1 keV interstellar cutoff
(but only because of our ability to discern the PWN's true shape from
the hard band image). The PWN is clearly extended.  It is also highly
asymmetric, with significantly greater extent to the north, opposite
the pulsar's apparent projected direction of motion. The spectra of 
the pulsar and PWN
can be described as power laws. There is no evidence for a $\sim$10$^6$~K
thermal component that might be associated with surface emission
from the pulsar.  Such emission is presumably not detectable as a
consequence of the background thermal emission from W44 and/or the
high column density. The nebular spectrum is softer than the pulsar,
and hints at steepening with distance from the pulsar.

The X-ray morphology of the PWN dif
fers crucially from the radio
morphology in two ways.
First, it is a factor of two smaller in extent along the
pulsar direction of motion. This difference can be ascribed to the 
overall softening of the
electron spectrum as the higher energy electrons lose their energy more
rapidly via synchrotron radiation. Second, the X-ray surface brightness peaks
at or very near the pulsar, and decreases monotonically with distance
from it. In contrast, the PWN has a more complex radio structure.  The
pulsar is embedded in a neck of emission 10$\arcsec$ wide that
connects with the larger cometary nebula $\sim$15$\arcsec$ to the north.
There is little diffuse radio emission surrounding the pulsar itself.
A bridge of emission connects the pulsar and the PWN to its north.
The radio surface brightness peaks not at the pulsar but near the
northern edge of the X-ray nebula.  It is unusual to see such
pronounced structural differences in a PWN.  More usually the X-ray
and radio surface brightness profiles resemble each other
and peak near the pulsar.  It is possible that low surface brightness 
structure associated
with the PWN in both the radio and the X-ray is lost as a consequence
of low contrast with the surrounding emission from the thermal remnant.

Is is surprising that any extent of the X-ray nebula is found. 
Based on synchrotron lifetime arguments,
electrons sufficiently energetic to produce X-rays were expected only
in close proximity to the pulsar (Harrus et al. 1996). In principle, the
combination of the X-ray and radio profiles along the direction of
the pulsar motion provides a record of the history of
the production of energetic particles in the pulsar magnetosphere.

 \begin{deluxetable}{rccccc}
 \tablecolumns{6}
 \tabletypesize{\footnotesize}
 \tablecaption{Parameters for Various Pulsar Wind Nebulae}
 \label{tab:pwn}
 \tablewidth{0pt}
 \tablehead{
 \colhead{} &
 \colhead{3C 58} &
 \colhead{Crab} &
 \colhead{G21.5-0.9} &
 \colhead{W44} &
 \colhead{IC 443}}
 \startdata
 X-ray Size (pc) & 10x6 & 1.2 & 7 & 1x0.5  & 3.5x2 \\
 Radio Size (pc) & 10x6 & 3.5x2.3 & 2.2x1.3 & 2x1 & 1.3x0.9 \\
 Distance (kpc) & 3.2 & 2 & 5 & 2.6 & 1.5 \\
 Age (yr)       & 820 & 950 & 3-6000 & 20,000 & 30,000 \\
 L$_x$ (ergs s$^{-1}$) & 2.4$\times$10$^{34}$ &  2.1$\times$10$^{37}$ &
   3.3$\times$10$^{35}$ &  6.0$\times$10$^{32}$ &
   2.6$\times$10$^{33}$ \\
 \.{E} (ergs s$^{-1}$) & 4.0$\times$10$^{36}$ &  4.7$\times$10$^{38}$ &
   3-6$\times$10$^{37}$ &  4.3$\times$10$^{35}$ &
   1.3$\times$10$^{36}$ \\
 L$_x$/\.{E} & 0.006 & 0.05 & 0.005-0.01& 0.001 & 0.002 \\
 $\Gamma$ range & 1.85, 2.4 & 1.6-2.3 & 1.5-2.8 & 2.1-2.3 & 1.6-2.3 \\
 \.{E} (ergs s$^{-1}$) & 4.0$\times$10$^{36}$ &  4.7$\times$10$^{38}$ &
   3-6$\times$10$^{37}$ &  4.3$\times$10$^{35}$ &
   1.3$\times$10$^{36}$ \\
 $\sigma$ & 2-15$\times$10$^{-3}$ & 3$\times$10$^{-3}$ &
  0.4-1.1$\times$10$^{-3}$ & 0.4-1.0$\times$10$^{-3}$ & \\
 Cutoff frequency (Hz) & 5$\times$10$^{10}$ & 1$\times$10$^{13}$ & &
  8$\times$10$^{12}$ & 1$\times$10$^{11}$ \\
 References\tablenotemark{a}  & 1,2 & 3,4 & 5 & 6,7 & 8 \\
 \enddata
 \tablenotetext{a}{REFERENCES -- (1) Torii et al. 2000; (2) Bocchino
 et al. 2001; (3) Kennel
 \& Coroniti 1984a; (4) Willingale et al. 2001; (5) Safi-Harb et
 al. 2001; (6) Wolszczan, Cordes \& Dewey 1991; (7) Frail et
 al. 1996; (8) Bocchino \& Bychkov 2001.}
 \end{deluxetable}

In Table 2, we compare the values of the various parameters for the
W44 PWN with those of other PWNs, including the young Crab and 3C~58,
the intermediate age G21.5-0.9, and the older IC~443 PWN.  In the
discussion that follows, we describe how the values for the W44 parameters
were inferred, and compare them with the values from the other objects.

The X-ray flux in the 2.2-8.2 keV band, where the spectrum was measured,
is 2.7$\times$10$^{-13}$~ergs~cm$^{-2}$~s$^{-1}$. This is considerably
lower than that
inferred by Harrus et al. (1996), but it can be expected that the
considerably greater difficulty of extracting the PWN signal from
the lower angular resolution ASCA data led to a less accurate flux
estimate. 
(Note that the ASCA detection is barely significant.) Extrapolation
of the unabsorbed flux to the Einstein band (0.2-4.0 keV) yields
a value of 7$\times$10$^{-13}$~ergs~cm$^{-2}$~s$^{-1}$, and a
corresponding luminosity of
6$\times$10$^{32}$ d$_{2.6}{^2}$ ergs~s$^{-1}$ (40,000 times less
luminous than the Crab).
This may be compared with the luminosity of
8$\times$10$^{32}$~ergs~s$^{-1}$ predicted using the empirically 
derived relation for pulsar wind nebulae 
(Seward \& Wang 1988)
log(L$_x$)=1.39log(\.{E})-16.6, where L$_x$ is the PWN X-ray luminosity
and 
 \.{E} is the rate of rotational energy loss by 
the pulsar, which has a value of 4.3$\times$10$^{35}$~erg s$^{-1}$
(Wolszczan, Cordes \& Dewey 1991).
The correspondence is remarkable, considering the nebula's atypical
morphology and history.

The spectral index is not observed to change radically with distance
from the pulsar, in contrast to all the other PWNs listed in Table 2.  
Most interesting is the contrast with the
most similar object known, the PWN in IC~443, whose spectral index
varies by 0.7 (Bocchino \& Bychkov
2001). As pointed out above, the overall nebular
X-ray spectral index is also similar to that of the Crab Nebula.
Chevalier (2000) developed a model for the X-ray luminosity of
PWNs in which he claimed that the X-ray spectral index is an indicator of
the efficiency with which the particle energy is converted into X-ray
emission.  In particular, he argues that in PWNs with Crab-like X-ray
(and thus electron) spectra, the X-ray luminosity should be produced
with high efficiency.  For simplicity, his model assumes a constant
magnetic field.  The field in the W44 PWN might be nearly constant,
given the lack of X-ray spectral index
variation. According to this model, the value and constancy of the X-ray
spectral index suggest that the W44 PWN
should have an \.{E}/L$_x$ that is much closer to the Crab than the
observed factor-of-five difference.  Since the only other PWN that
seems Crab-like in this regard according to Chevalier (the LMC remnant
0540-69.3) is, like the Crab, about 1,000 yr old, perhaps some additional
factor, such as the age of the nebula, needs to be incorporated in the model.

In the widely accepted model of PWNs, the energy source is the 
spindown energy of the pulsar.  The energy is efficiently 
transferred into a relativistic wind with some characteristic Lorentz factor
$\gamma$.  This energy is divided between Poynting (magnetic and 
electric) flux and particle energy flux, the ratio of which is the 
parameter $\sigma$. The initial interaction between the wind and the 
surrounding medium forms a termination shock, usually creating wisps and 
filaments identifiable in the radio and visible bands (and now in the 
X-ray, using Chandra).  In the pulsar's magnetosphere, where the
particles are created by pair production, the wind is 
expected to be magnetically dominated ($\sigma$~$\ge$~1).  
By the time the wind reaches the termination shock, 
observations indicate it becomes particle dominated 
($\sigma$~$<$~1).  
No clear theoretical explanation for this transition has emerged
(Arons 1998).
Measurements of wind parameters in other PWNs
typically show $\sigma$ on the order of a few times 10$^{-3}$
(see Table 2).  The composition of the particle flux is 
also of interest.  This is characterized by the parameter k, the 
ratio between the energy density in electrons, and that in all other 
particles (positrons, protons and other nuclei).  While models 
commonly assume an electron-positron plasma with k~=~1, Frail et al. 
(1996) have shown that in the W44 wind nebula 5~$\le$~k~$\le$~30.  

The Chandra observation allows us to refine estimates of
these parameters characterizing the PWN, as well as the magnetic 
field strength B. Under the assumption
that the break between the radio and the X-ray spectra arises from
synchrotron losses, Frail et al. (1996) have shown how an estimate
of the turnover frequency, $\nu_B$, of the PWN spectrum leads to estimates
of the nebular magnetic field strength and the Lorentz factor, $\gamma$,
for the electrons near $\nu_B$. From the improved measurement of the
X-ray flux and spectral index of the PWN, we find a best-fit cutoff
frequency $\nu_B$ = 8$\times$10$^{12}$~Hz. The reduced flux value is 
the primary
reason why the estimate of the synchrotron cutoff in the present
work is substantially lower than that of Harrus et al. (1996).
Our best-fit $\nu_B$ is similar to
the Crab's break frequency of 10$^{13}$~Hz. Using the equations
reproduced by Frail et al. from Pacholczyk (1970) we then find:

  B = 1040 $\mu$G ($\nu_B$/10$^{12}$ Hz)$^{-1/3}$(t$_{res}$/1000 
yr)$^{-2/3}$ = 160 $\mu$G; and 

  $\gamma$ $\sim$ 10$^5$ ($\nu_B$/10$^{12}$ Hz)$^{1/2}$ (B/100 
$\mu$G)$^{-1/2}$ = 2.2$\times$10$^5$

\noindent
Here t$_{res}$ represents the age of the nebula, which Frail et al. 
take to be 5,700 yr based on the nebular extent in the radio.

The new cutoff frequency estimate falls squarely between the values
used by Frail et al. (1996) to bound the value of k.
Using our best value of $\nu_B$, we find k $\sim$ 10, 
consistent with their conclusionthat the particle energy is 
electron dominated.

We can estimate the value of the magnetization parameter
$\sigma$, following the approach used by
Torii et al. (2000) for 3C~58. They used the
formalism developed by Kennel and Coroniti (1984a, b) for the Crab,
who showed that $\sigma$ is related to the velocity profile
of the nebula by v(z)/c~$\sim$~3$\sigma$[1~$+$~(3z$^2$)$^{-1/3}$].  Here
z = r/r$_s$, the ratio between radius and the distance between the
pulsar and the termination shock. Torii et al. estimated $\sigma$ by
es
timating r$_N$ and v$_N$, the values at the edge of the radio
nebula. Unlike 3C 58, in which the pulsar shows little proper
motion, care must be exercised here in estimating global nebular
parameters like size in the presence of substantial pulsar motion.

For the
nebular size r$_N$ we take the largest dimension 
in the radio nebula perpendicular to
the direction of motion, which is 1 arc minute or 0.75~d$_{2.6}$~pc.
For the nebular age, we cannot use the nominal spin down age
of 20,000~yr. Given the estimated proper motion, the pulsar has
moved $\sim$8 arc minutes, an angular distance far larger than the
size of the observed nebula. 
Instead we use the age of the radio nebula, 5,700 yr, which is based
on synchrotron lifetime arguments (Frail et al. 1996).
Assuming homologous expansion, we find
v$_N$ (= 2/5 $\times$ 0.75~pc/5,700~yr)~$\sim$~50~km~s$^{-1}$; assuming 
constant expansion, we find v$_N$~$\sim$~130~km~s$^{-1}$.
Another estimate of the nebular lifetime can be derived 
from the time required by the pulsar to move from the location
corresponding to the largest perpendicular dimension
to its current location;
1.5 arc minutes divided by the
proper angular motion of 25 mas/yr, or 3,600 yr.
This simple calculation suggests that the lifetime used
can be off by not more than a factor of $\sim1.6$.

There is no
evidence in the visible, radio or X-rays of the wisps observed
in the Crab or 3C~58 that are interpreted as the termination shock.
Since the termination shock represents the location where the ram
pressure of the relativistic wind from the pulsar (\.{E}/(4$\pi$cr))
equals the pressure in the nebula (presumably dominated by magnetic
pressure), we can estimate r$_s$ by equating these two quantities at
the termination shock. 
Thus \.{E}/(4$\pi$c r$_s$)=1/3(B$^2$/(8$\pi$)).  For B=160 $\mu$G and 
\.{E}=4.3$\times$10$^{35}$~ergs~s$^{-1}$ (Wolszczan, Cordes \& Dewey
1991), we find r$_s$~$\sim$~0.02~pc, leading to a value of
z$_N$~=~r$_N$/r$_s$~$\sim$~80, which can be compared with z$_N$ of 20 for 
the much younger and more energetic Crab, and 15-100 for 3C~58
(Torii et al. 2000).

Combining the estimates for z$_N$ and v$_N$,
we estimate $\sigma~\sim~$0.4-1.0$\times$10$^{-3}$, depending upon which value of
v$_N$ is assumed. As indicated in Table 2, this value of $\sigma$ is 
lower than
that found for the younger PWNs, including the 
Crab (3$\times$10$^{-3}$ -- Kennel
\& Coroniti 1984a), 
3C~58 (2-15$\times$10$^{-3}$ -- Torii et
al. 2000), but is consistent with that for the older
G21.5-0.9 (4-11$\times$10$^{-4}$ Safi-Harb et al. 2001).
While the sample of measurements is small, it suggests that 
$\sigma$ may decrease with PWN age, but it either stays near a value of 
$\sim$10$^{-3}$ over many thousands of years or $\sigma$ must have a
value around 10$^{-3}$ if a PWN is to be observable.

We can use the refined magnetic field estimate to obtain a 
synchrotron lifetime estimate for the 
 hard band X-ray
emission pictured in Figs. 1 and 2. 
For a magnetic field strength B in Gauss, and a photon frequency
$\nu$ in Hz, the synchrotron lifetime $\tau$ in seconds is
$\sim$6$\times$10$^{11}$B$^{-3/2}\nu^{-1/2}$.  For the estimated
value of B (160 $\mu$G) and a photon frequency of 5$\times$10$^{17}$~Hz 
($\sim$2.5~keV), we find
$\tau~\sim$3$\times$10$^8$~s, or approximately 15 yr.  
This lifetime is very short compared
with the $\sim$20,000~yr pulsar age (Wolszczan, Cordes \& Dewey
1991) or the apparent nebular age of 5,700 yr. 
The short lifetime, along with the $\sim$1~d$_3$~pc extent of the
X-ray nebula implies that the X-ray emitting
electrons ($\nu$~$\ge$~5$\times$10$^{17}$~Hz) have a high streaming velocity, 
$\sim$1/3~c. The high streaming velocity in
turn suggests that the magnetic field in the extended nebula is ordered,
and oriented along the wake of the pulsar's motion. 
It is reasonable to expect an ordered field if one considers
that the PWN magnetic field should be considerably stronger than any residual
magnetic field in the remnant interior, even in the mixed-morphology
W44 whose interior density is substantially higher than a typical
shell-like SNR.

As a final note, one feature of potential future
interest is the possible low surface brightness extended X-ray
emission. If it is nonthermal emission associated with the PWN, 
then W44 would be the third PWN with a larger apparent extent in the 
X-ray than in the radio.
Similar structures have been
found in the plerionic remnant G21.5-0.9 (Slane et al. 2000;
Warwick et al. 2001) and the PWN inside IC~443 (Bocchino \& Bychkov
2001), with clearly nonthermal X-ray emission
extending well beyond the radio nebula. If the emission in W44
is confirmed, it is possible that low surface brightness radio
emission is also produced but is invisible against the foreground
emission from the shocked gas in the supernova remnant.
The existence of such low level
emission has been speculated upon by Warwick et al. (2001) for
G21.5-0.9, where it might be easier to detect. The reality of the
X-ray emission should be straightforwardly demonstrated by
XMM/Newton, whose substantially higher throughput will facilitate
more accurate measurements of the PWN spectrum and its variations.

In summary, the ACIS observation of the pulsar wind nebula 
surrounding PSR~B1853+01 in W44 reveals an extended nebula, half the 
size of the radio PWN.
Spectroscopy reveals a significant difference between the power law 
photon index of PSR~B1853+01 ($\Gamma$~$\sim$~1.4) and that of the 
nebula ($\Gamma$~$\sim$~2.2).  Variation of the photon index within 
the nebula has not been detected.  The X-ray size and spectrum of the 
PWN have allowed us to estimate key parameters, including magnetic 
field strength, the average $\gamma$ of the particles in the wind, 
the magnetization parameter ($\sigma$), and the ratio k of electrons 
to other particles.  
We find that despite the unusual morphology produced by the high
velocity and age of the pulsar, the W44 PWN has properties similar
typical of other PWN. 
A number of unresolved issues, such as the possible existence an
extended nebula and spectral variation within the nebula, make this
fascinating object deserving of more extensive study.
It is
hoped that deeper observations using either Chandra or XMM/Newton can 
provide
more exact values of the observables and facilitate measurements of 
their variation with distance from PSR~B1853+01, leading in turn to 
more robust estimates of the PWN parameters.

\acknowledgments
We acknowledge the tireless efforts of the CXC staff, and the team
that developed the Chandra observatory.
We would like to thank D. Frail for use of the 20~cm map,
and K. Arnaud with assistance while we were developing
the IDL software used to extract irregular regions.
Support for this work for R.L.S. and K.D.K. was provided by the 
National Aeronautics and Space Administration through grant G01-2057A 
issued by the Chandra X-ray Observatory for and on behalf of NASA 
under contract NAS8-39073.

\clearpage


%

\end{document}